\documentclass[a4paper,twocolumn]{esapub} 
          [2001/04/25 1.1 (PWD)]
\usepackage{times}
\usepackage{natbib}
\usepackage{graphics}
\usepackage{amssymb} 
\title{Effect of local treatments of convection upon the solar p-mode excitation rates}
\author{R\'eza Samadi}
\author{M.J. Goupil}
\author{Y. Lebreton}
\author{C. Van't Veer}
\affil{Observatoire de Paris, 5 place J. Janssen, 92195 Meudon, France}
\author{F. Kupka}
\affil{Max-Planck-Institute for Astrophysics, Karl-Schwarzschild Str. 1, 85741 Garching}

\newcommand{\apj} {ApJ}
\newcommand{\mnras} {MNRAS}
\newcommand{\aap} {A\&A}

\newcommand{\aaps} {A\&AS}

\newcommand{\fig}[3]{
      \begin{figure}[tbp]
	\resizebox{\hsize}{!}{\includegraphics  {#1}}
	\caption{#2}
	\label{#3}
        \end{figure} }
\newcommand{\eqn} [1] {
\begin{equation} 
#1 
\end{equation}}

\begin{document}


\maketitle

\begin{abstract}
We  compute, for several solar models, the  rates $P$ at which the  solar  radial $p$~modes  are expected to be excited.
The solar models are computed with two different  local treatments of convection  :
the classical mixing-length theory (MLT hereafter)  and  \citet[][CGM hereafter]{Canuto96}'s formulation.

For one set of solar models (EMLT  and ECGM models), the atmosphere is gray and assumes  Eddington's approximation. The   models  only  assume  one  mixing-length parameter and reproduce the  solar radius  at solar age but not the Balmer line profiles.
For a second set of models (KMLT  and KCGM models), the atmosphere  is built using a $T(\tau)$ law which has been obtained  from a Kurucz's  model  atmosphere computed with the same local treatment of convection.  
The   mixing-length parameter  in  the model atmosphere  is chosen  
so as to provide a good agreement  between synthetic and observed Balmer line profiles, while the mixing-length parameter  in the interior model  is calibrated so that the model  reproduces the solar radius  at solar age.  

For the MLT treatment, the rates $P$   do depend significantly on  the properties of the  atmosphere.
Indeed differences in $P$ between the EMLT model and the KMLT are found very large.
On the other hand, for  the CGM treatment,  differences in $P$ between the ECGM  and the KCGM models are very small compared to the error bars attached to the seismic measurements.
The  excitation rates $P$ for modes from the EMLT model  are significantly under-estimated  compared with the solar seismic constraints.
The KMLT model  results in intermediate values for $P$ and shows also an important discontinuity in the temperature gradient and the convective velocity.
On the other hand,  the KCGM model and the ECGM model yield values for $P$  closer to the seismic data than the EMLT and KMLT models.
We conclude that the solar p-mode excitation rates  provide valuable constraints and according to the present investigation cleary favor the CGM treatment with respect to the MLT.
\end{abstract}

\section{Introduction}
We study the implication of two different local treatments of convection on the calculation of the rates P at which solar p modes are excited by turbulent convection. For this purpose, solar models  are built with two different  local treatments of convection  : the classical MLT  and  \citet[][]{Canuto96}'s formulation (Sect.~\ref{solar_models}).

For a first set of models (EMLT  and ECGM models), the atmosphere is gray and assumes  Eddington's approximation. Each of these  models  therefore   uses  one  single value of the mixing-length parameter and reproduces the  solar radius  at solar age but not the Balmer lines.

For a second set of solar models (KMLT  and KCGM models), the atmosphere is built using a $T(\tau)$ law which has been obtained  from Kurucz's  model  atmospheres computed with the same local treatment of convection as used in the interior.  
The   mixing-length parameter  in  the model atmosphere ($\alpha_a$)  is chosen  so as to provide a good agreement  between synthetic and observed Balmer line profiles, while the mixing-length parameter  in the interior model  ($\alpha_i$) is calibrated so that the model  reproduces the solar radius  at solar age.

Calculation of the excitation rates requires  the computation of the convective velocity and the convective flux.  This is done in Sect.~\ref{fc} and \ref{v} by paying  special attention to the difficulty of matching interior and atmosphere models.
We finally present the computed  excitation rates P  for each model  and compare with the solar data in Sect.~\ref{results}.

\section{Calculation of the rate at which solar p-modes are excited}
\label{P_calculation}

The model of stochastic excitation (MSE)  used here  is basically that of \citet[][ see also Samadi et al. 2004, present conference]{Samadi00I}. This model provides an expression for $P(\nu)$, the rate at which a given radial mode with frequency $\nu_0$  is excited and requires a proper description of the structure of the very outer layers.

In practice the calculation of $P$ needs the following quantities as input:
1) The mean density,  the  convective velocity ($v$) and the  entropy fluctuations ($s$) of the solar model. $s$ is directly proportional to  the ratio $(F_{\rm c}/v)^2$ where $F_{\rm c}$ is the convective flux. $F_{\rm c}$ and $v$ are calculated as explained in Sect.~\ref{P_calculation}.
2) The eigenfunctions ($\xi$) and their frequencies ($\nu_0$). They are computed with the adiabatic code FILOU of \citet{Tran95}.
$E(k)$ the wavenumber ($k$) dependency of the turbulent kinetic energy spectrum, $E(k,\nu)$.
3) The  values and depth dependence of the wavenumbers $k_0$, the wavenumber at which $E(k)$ is maximum. 4)  $\chi_k (\nu)$, the frequency ($\nu$)  component of  $E(k,\nu)$

According to the results in \citet{Samadi02I,Samadi02II}   obtained on the base of a solar 3D simulation
 the $k$-dependency of $E(k,\nu)$ is approximatively reproduced by an analytical spectrum called  'Extended Kolmogorov Spectrum' (EKS) and defined in \citet{Musielak94}. 
 The $\nu$-dependency of $\chi_k$ is found to be  better modelled with a Lorentzian function rather than by a Gaussian function which is usually assumed for $\chi_k$  \citep[see ][ present conference]{Samadi04b}. 
 At the top of the superadiabatic region,  it was found that  $k_0 \sim 3.6 ~{\rm Mm}^{-1}$  and decreases slowly with depth inward. In this work we assume a constant $k_0 = 3.6 ~{\rm Mm}^{-1}$  which is a good approximation.

\section{The solar models}
\label{solar_models}

The models are computed with the CESAM code  \citep{Morel97}
 including the following input physicses:
CEFF  equation of state \citep{JCD88}, OPAL opacities \citep{Iglesias96} data complemented by \citet{Alexander94} data for $T\lesssim 10^4\ \rm K$, both set of data being given for \citet{GN93} solar mixture,  \citet{Caughlan88} thermonuclear reaction rates,  microscopic diffusion  according to the simplified formalism of  \citet{Michaud93} and finally 
  B\"ohm-Vitense formulation of the MLT \citep{Bohm58}  or CGM's formulation (the same formulation in the interior as in the atmosphere).

{\it ECGM and EMLT models:}
The atmosphere of the ECGM and EMLT is  gray and assumes  Eddington's approximation.
The mixing-length parameter $\alpha$ of those models (the same $\alpha$  in the interior as in the atmosphere)  is adjusted in order to reproduce the solar luminosity and radius at the solar age.
These models do not reproduce the Balmer line profile.
 Table~\ref{tab:edd} gives the calibrated values of the mixing-length parameters.

\begin{table}[b]
\caption{Values of the the mixing-length parameter $\alpha$ of the ECGM and EMLT models.
}
\label{tab:edd}
\begin{center}
\begin{tabular}{cccc}  
model &$\alpha$ & \cr
\hline
EMLT & 1.763     \cr
ECGM &0.91      \cr
\hline 
\end{tabular}
\end{center}
\end{table}

{\it KCGM and KMLT models}:
The atmosphere of the KCGM and KMLT are  restored from a 
$T(\tau)$-law derived from Kurucz's complete ATLAS9 model atmospheres according to the
procedure described in \citet{Morel94}. 
The fit between the interior (where diffusion approximation is valid) and atmosphere is performed in the region
where $\tau_1 < \tau < \tau2$.  In the interior region where $\tau  > \tau_2$, 
the temperature gradient $\nabla_{\rm i}$ is obtained from the MLT or CGM formulation. In the atmospheric region where $\tau < \tau_1$, the temperature gradient $\nabla_{\rm a}$ 
is computed using the $T-\tau$ law of the model atmosphere built with the same formulation of 
convection as in the interior. In the transition region where $\tau_1 < \tau < \tau_2$, in order to ensure
the continuity of the temperature gradient, $\nabla$ is obtained
by a linear interpolation of $\nabla_{\rm i}$ and $\nabla_{\rm a}$ as a function of the optical depth $\tau$, as follows:
\eqn{
\nabla = \beta(\tau) \nabla_{\rm a} + (1-\beta(\tau)) \nabla_{\rm i}
\label{eqn:nabla:full_model}
}
where $\beta(\tau) =\displaystyle  (\tau_2-\tau)/(\tau_2-\tau_1)$.
In practice we found that $\tau_1$=5 is the minimal acceptable value  for $\tau_1$ and $\tau_2$=20 minimizes the discontinuity between the interior and the atmosphere.

{\it Calibration of the KCGM and KMLT models}:
 The mixing-length parameter $\alpha_a$ of the  model atmosphere is adjusted so as to provide the best agreement between synthetic and observed Balmer line profiles \citep[as in ][ for the MLT treatment]{Vantveer96}. For all the treatments of convection the value  $\alpha_a=0.5$ provides the best agreement between synthetic and observed Balmer line profiles.
 The mixing-length parameter $\alpha_i$ of the internal structure is adjusted in order that the global model reproduces simultaneously  the solar radius and luminosity at the solar age. 
Values of the calibrated mixing-length parameters are given in Table~\ref{tab:kurucz}.
\begin{table}[b]
\caption{Values of the the mixing-length parameters of the KCGM and KMLT models: 
$\alpha_{\rm i}$ (for the interior) and  $\alpha_{\rm a}$ (for the model atmosphere). 
}
\label{tab:kurucz}
\begin{center}
\begin{tabular}{cccc}  
model  &$\alpha_{\rm i}$ & $\alpha_{\rm a}$\cr
\hline
KMLT & 2.42  & 0.50 \cr
KCGM & 0.78 & 0.50 \cr 
\hline
\end{tabular}
\end{center}
\end{table}

\subsection{Calculation of the convective flux}
\label{fc}

One contribution to the driving of the oscillation modes comes from the advection of turbulent entropy fluctuations by the turbulent movements (the so-called entropy source term). It scales as the square of the convective flux $F_{\rm c}$. 

$F_{\rm c}$ can be viewed as function of $\alpha$ and $\nabla$: $F_{\rm c}=h(\nabla ,\alpha )$. $h$ is given by the adopted formulation of convection (here MLT or CGM).
In the outer region ($\tau < \tau_1$) and in the interior region ($\tau > \tau_2$) $F_{\rm c}$  is computed directly from the function $h$.
In the transition region, as in Eq.~(1), the convective flux $F_{\rm c}$ can be related to the convective flux of the interior ($F_{\rm c}^{\rm i}$) and the convective flux of the atmosphere ($F_{\rm c}^{\rm a}$) as follows:
\eqn{
F_{\rm c} = \lambda(\tau) \, F_{\rm c}^{\rm a} + (1-\lambda(\tau)) \, F_{\rm c}^{\rm i}
\label{eqn:fc:full_model}
}
In Eq.~(2), $F_{\rm c}^{\rm i}=h(\nabla_{\rm i},\alpha_{\rm i})$, $F_{\rm c}^{\rm a}=h(\nabla_{\rm a},\alpha_a)$, $\lambda (\tau)$  is,  as $\beta (\tau)$  (Eq.~1), a function to  ensure the  continuity of the convective flux. The choice for $\lambda (\tau)$, as for $\beta (\tau)$, is rather arbitrary. We assume $\lambda (\tau)=\beta (\tau)$  for sake of simplicity. Results of the calculation of $F_{\rm c}(\tau)$ are shown in Fig.~(1).

\fig{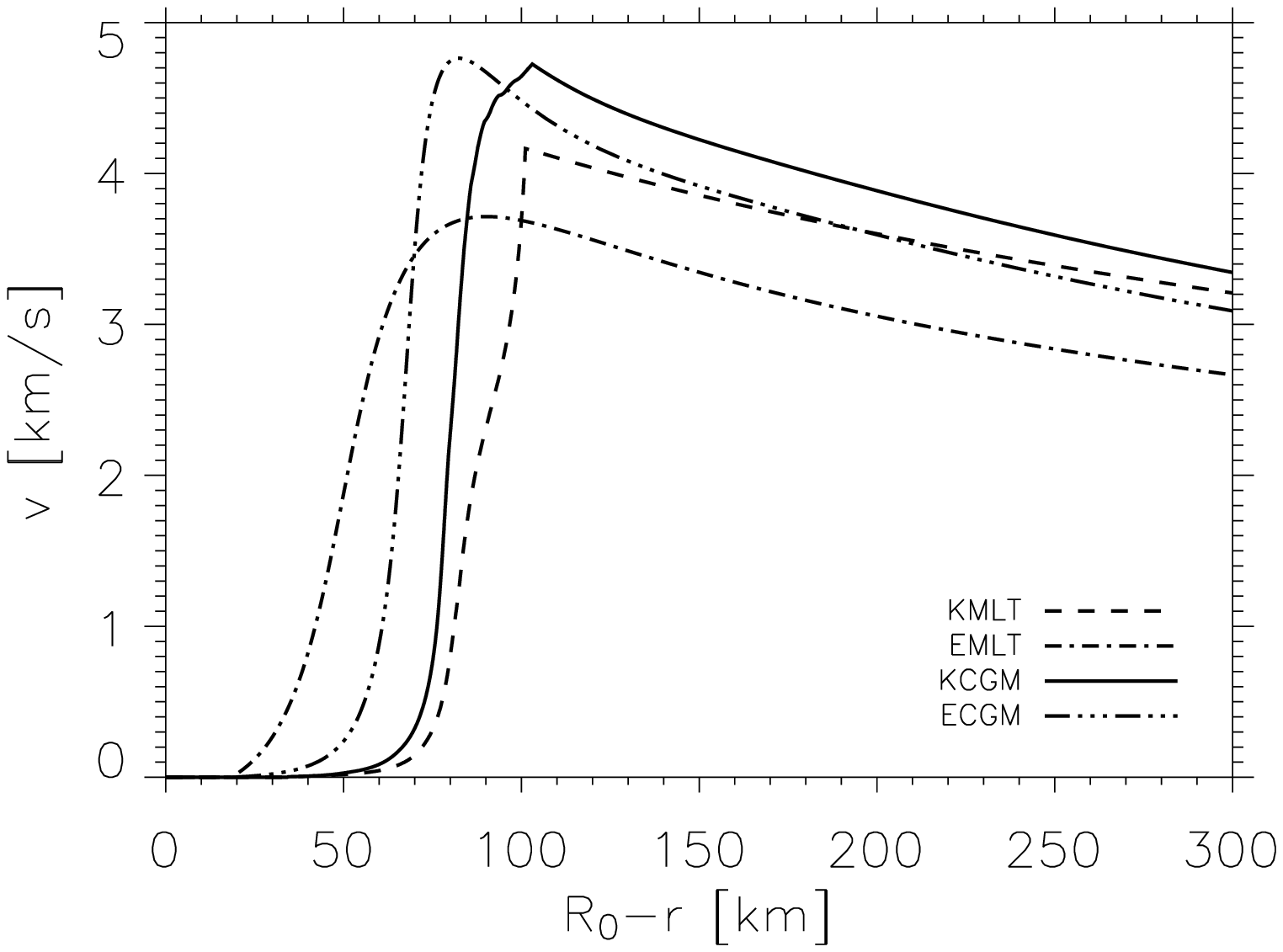}{The computed convective $F_{\rm c}$ is plotted  versus the optical depth $\tau$  for the KMLT model (dashed line), the EMLT model (dot dashed line), the KCGM model (solid line) and the ECGM model (dot dot dashed line). }{fig:1}

\subsection{Calculation of the convective velocity}
\label{v}
The second contribution to the driving of the oscillations comes from the Reynolds term and is proportionnal to $v^4$ where $v$ is the convective velocity.

The convective velocity ($v$) can be viewed as a function of $\nabla$ and $\alpha$: $v=f(\nabla,\alpha)$. $f$ is given by the adopted formulation of convection (here MLT or CGM).

For each model, in the outer region (i.e.  $\tau < \tau_1$) and in the inner region ($\tau >\tau_2$), given $\nabla$ and  $\alpha$, $v$ is  computed as $v=f(\nabla,\alpha)$.
In the transition region ($\tau_1< \tau <\tau_2$) of the KMLT and KCGM models we must deal with two different values of $\alpha$ (namely $\alpha_a$, $\alpha_i$) but $v$ is not a linear function of $\alpha$. We thus face the difficulty to properly define a convective velocity consistent with $F_{\rm c}$ in this  region (Eq.~2). 

We then proceed as follows: $\nabla$   and  $F_{\rm c}$ are  defined by Eq.~(1) and Eq.~(2) respectively. Then at fixed $\nabla$ and $\tau$, we define  an equivalent mixing-length parameter, $\alpha^*$, such that $F_{\rm c}= h(\nabla,\alpha^*)$. Such mixing-length parameter is hence variable with depth. We next compute  $v=f(\nabla(\tau),\alpha^*(\tau))$ which then is consistent with $F_{\rm c}$ of Eq.~(2).

\fig{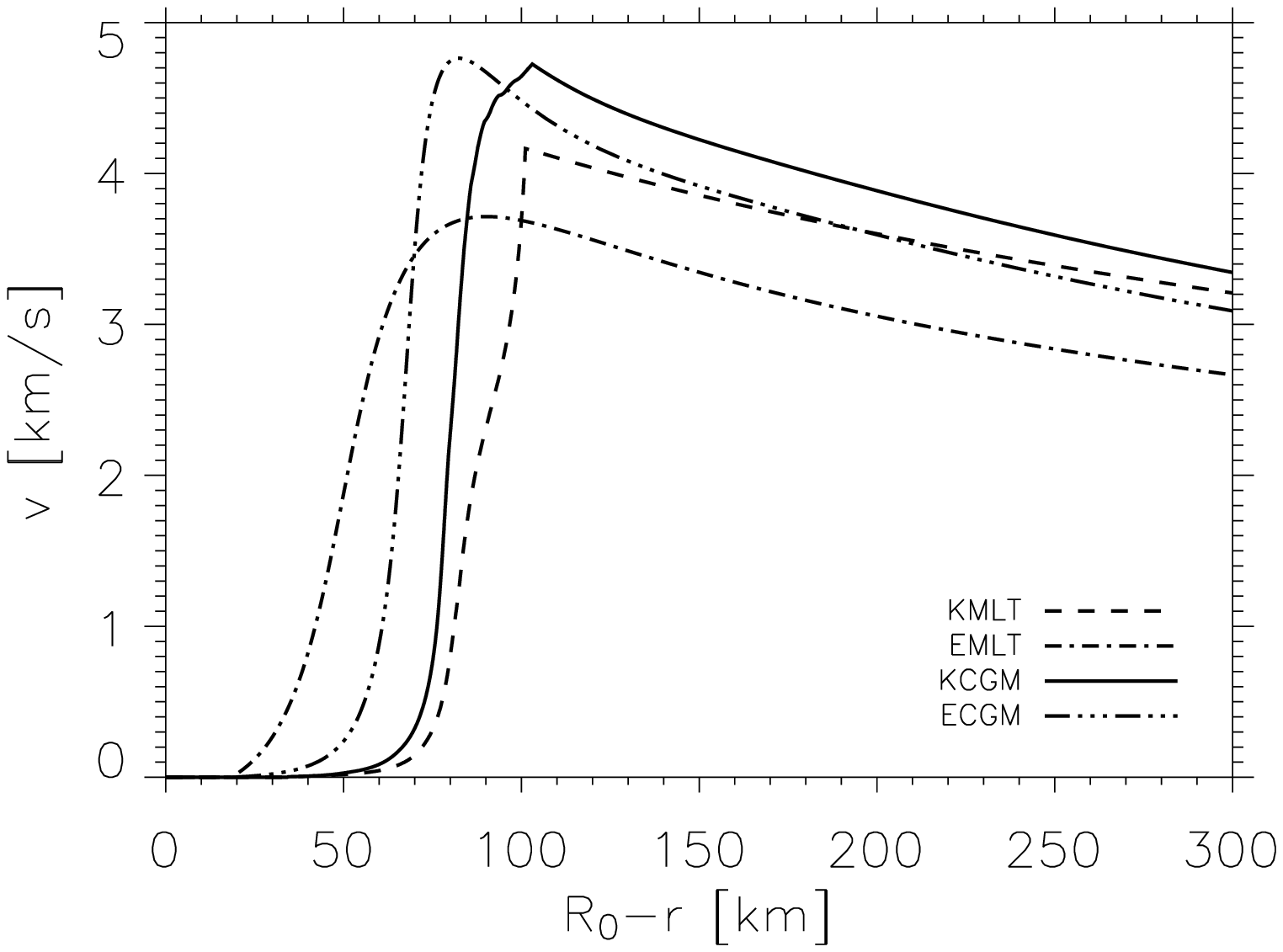}{The computed convective velocity $v$ is plotted versus $R_0-r$ where $R_0$ is the solar radius and $r$ the radius of a given layer. The legend has  the same meaning as in Fig.~(1).}{fig:2}

\subsection{Comments}

The CGM model results in a much smaller value for $\alpha_i$ than the MLT model. Furthermore for the CGM model,  $\alpha_i$  is found closer to $\alpha_a=0.5$. In contrast, for the MLT treatment, the value of  $\alpha_i$ is much larger than $\alpha_a=0.5$. These features can be explained by the fact that in nearly adiabatic regions convection is more  efficient in the CGM's formulation (see details in Heiter et al, 2002). 
 Differences in $\alpha_i$ between the KGCM and the ECGM models are relatively smaller than between the KMLT and the EMLT models.

As shown in Fig.~(1) and (2), the EMLT and KMLT models have very different convective structures. This is due to large differencies between $\alpha_i$ and $\alpha_a$. 
 KCGM and ECGM models have very similar convective structures. This is a consequence of the fact that the KCGM model results in a value of $\alpha_i$ close to that required for the atmosphere ($\alpha_a=0.5$).

For the KMLT model, there is an important discontinuity at the bottom boundary of the transition region ({\it i.e.} at $\tau =20$ or $R_0-r = 100$~km), especially for $v$. In contrast for the KCGM the discontinuity is much less important. These features are again directly connected with the differences between between $\alpha_i$  and $\alpha_a$. 

{\it Conclusion}: both the KMLT and the KCGM models reproduce  the Balmer line profile and the solar radius and luminosity but the CGM model  models the transition between the region of high convective efficiency (the interior) and  the region of low efficiency (the atmosphere) in a much better way than the MLT treatment.

\section{Results and conclusion}
\label{results}
Results of the calculations of the excitation rates $P$ are presented in Fig.~(3). The excitation rates $P$ inferred from the observations by Chaplin et al (1998) are also represented for comparison.

\fig{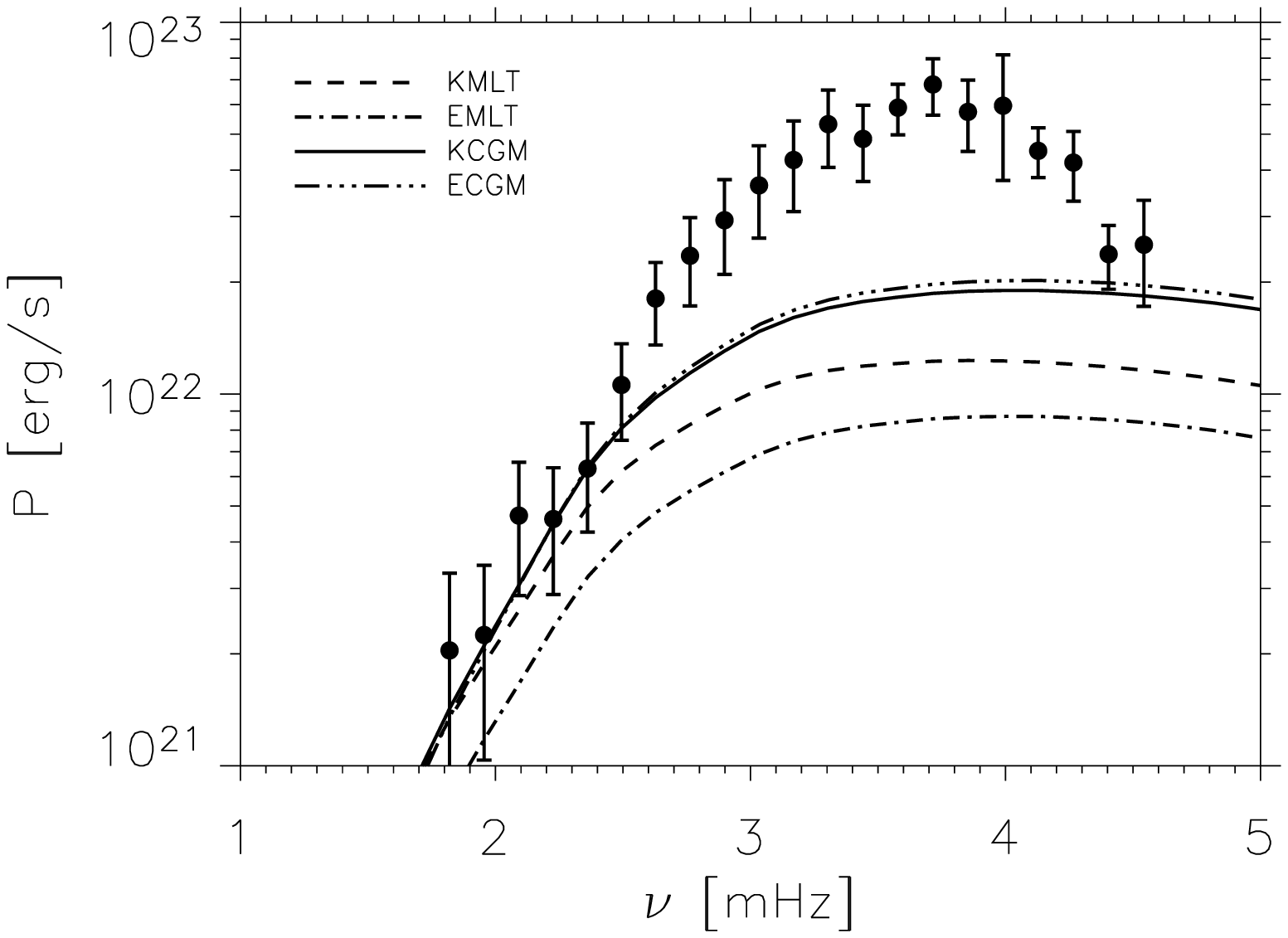}{Solar $p$~mode excitation rates, $P$, versus the mode frequencies n. Computed P are represented for the KMLT model (dashed line), the EMLT model (dot dashed line), the KCGM model (solid line) and the ECGM model (dot dot dashed line). The filled circles with associated error bars represent  the 'observed'  solar values of $P(\nu)$ derived from the amplitudes and line widths of the radial modes measured  by \citet{Chaplin98}.}{fig:3}

For the MLT treatment, the rates P   do significantly depend  on  the properties of the  atmosphere. Indeed differences in P between the EMLT model and the KMLT are found very large. On the other hand, for the 
 CGM treatment,  differences in P between the ECGM  and the KCGM models are very small compared to the error bars attached to the seismic measurements. 

For the  EMLT and KMLT models, P are significantly under-estimated  compared with the solar seismic constraints obtained from \citet{Chaplin98}'s measurements. 
 KCGM and the ECGM models yield values for P  closer to the  seismic data than the EMLT and KMLT models.  

We conclude that the solar p-mode excitation rates  provide valuable constraints on the properties of the super-adiabatic region. According to the present investigation (focused on local approaches), they clearly favor the CGM treatment with respect to the MLT one.

The remaining discrepancy between computed and observed P (Fig. 3) is due to the local assumption in the convective treatment. Indeed \citet{Samadi02II} have succeeded in reproducing much better  the seismic constraints  by using constraints from a solar 3D simulation \citep[see][ present conference]{Samadi04b}.

\section*{Acknowledgments}

 RS acknowledges support by Comit\'e National Francais d'Astronomie and by the Scientific Council of Observatory of Paris.


\bibliographystyle{aa}

\end{document}